\documentclass{aa} %

\usepackage{graphicx}
\usepackage{txfonts}
\begin{document} 

   \title{No unique solution to the seismological problem of\\ standing kink MHD waves}

\titlerunning{No unique solution to the seismological problem of standing kink MHD waves}
\authorrunning{Arregui \& Goossens}

   \author{I.  Arregui\inst{1,2} \and M. Goossens\inst{3}}
   \institute{
   Instituto de Astrof\'{\i}sica de Canarias, E-38205 La Laguna, Tenerife, Spain\\
   \email{iarregui@iac.es}
  \and 
  Departmento de Astrof\'{\i}sica, Universidad de La Laguna, E-38206, La Laguna, Tenerife, Spain
  \and
Centre for mathematical Plasma Astrophysics, Department of Mathematics, KU Leuven Celestijnenlaan 200B bus 2400, B-3001 Leuven, Belgium\\\email{marcel.goossens@kuleuven.be}}

   \date{Received ; accepted}

 
 \abstract{The aim of this paper is to point out that the classic seismological problem using observations and theoretical expressions for the periods and damping times of transverse standing magnetohydrodynamic (MHD) waves in coronal loops is better referred to as a reduced seismological problem. `Reduced' emphasises the fact that only a small number of characteristic quantities of the equilibrium profiles can be determined. Reduced also implies that there is no unique solution to the full seismological problem. Even the reduced seismological problem does not allow a unique solution. Bayesian inference results support our mathematical arguments and offer insight into the relationship between the algebraic and the probabilistic inversions.}
 

   \maketitle

\section{Introduction}\label{intro}

Seismology using periods and damping times of transverse standing magnetohydrodynamic (MHD) waves in coronal loops has become a tool for obtaining estimates of physical quantities that are difficult to measure with other means. The first example of the use of the period of a standing MHD wave is by \cite{nakariakov00a} and \cite{nakariakov01}.  \cite{nakariakov01} used the classic approximate expression for the period $P$ derived from the phase speed $v_{\rm ph} = v_{\rm k}$, with $v_{\rm k}$ the kink speed, to obtain estimates of the magnetic field strength. This classic expression is obtained for an equilibrium model with a piece-wise constant density in the direction across a constant vertical magnetic field in the thin tube approximation. If the interest is solely in the period as a possible tool for seismology,  then the constant density model suffices for a first comparison between observation and theory.  

Observations show that these standing MHD waves are damped with damping times of the order of 3 to 10 periods \citep[see e.g.][and others]{aschwanden99,nakariakov99}. The term damping time is the time that is needed to reduce the amplitude of the wave by a factor $e$. It was realised that resonant absorption offers an explanation for this observed damping. For resonant absorption see for example \cite{ionson78,hollweg88,goossens92a,ruderman02}.  However, for resonant absorption to work the equilibrium has to be non-uniform in the cross-field direction. Equilibrium models with a non-uniform intermediate layer were introduced in the seismological investigations. In the non-uniform layer of thickness $l$ the density varies in a continuous manner from its internal value $\rho_{\rm i}$ to its external value $\rho_{\rm e}$. \cite{goossens02a} used observed values and an approximate expression for the exponential damping time $\tau_{\rm D}$ to compute the thickness of the non-uniform layer, $l/R$ with $R$ the radius of the loop, for 11 events.  \cite{arregui07a} (fully numerical) and \cite{goossens08a} (analytical) used observed values of the period $P$ and the exponential damping time $\tau_D$ to obtain estimates for the density contrast $\zeta = \rho_{\rm i} /\rho_{\rm e}$, the longitudinal Alfv\'{e}n travel time $\tau_{\rm Ai}$ and the inhomogeneity length scale $l/R$.  \cite{arregui07a} and \cite{goossens08a}  noted  that their seismological inversion problem had $\infty^1$ solutions. As a matter of fact both \cite{arregui07a} and \cite{goossens08a} were too optimistic. The seismological inversion has $\infty^2$ solutions but the unknown steepness of the density profile was removed from the inversion scheme by adopting a sinusoidal variation of the equilibrium density.

Resonant absorption is a damping mechanism that works for both standing and propagating waves. Observations by the COronal Multi-channel Polarimeter (CoMP) show that propagating MHD waves are ubiquitous  in the solar atmosphere \citep{tomczyk07}. This is in stark contrast to standing MHD waves that are rather rare since they need an energetic event as excitation. This  triggered interest in the spatial damping of propagating waves by resonant damping. \cite{terradas10} derived an analytical expression for the damping length and confirmed its accuracy by numerical simulations. \cite{verth10} used the theory developed by \cite{terradas10} to successfully explain the disparity between outward and inward wave power of the waves observed with CoMP \citep{tomczyk09}

The damping time and the damping length discussed so far are exponential as the theory adopts a model with wave variables undergoing an exponential decay in time in case of standing waves or in space in case of propagating waves.  This comes naturally since the observed damped signals were modelled as exponentially damped harmonic functions.  Numerical experiments on propagating waves suggest that the initial damping is better described by a Gaussian rather than an exponential decay.  Gaussian damping is first studied for spatial damping of propagating waves by \cite{pascoe12}.  Analytical theory for spatial damping of propagating waves is developed by \cite{hood13}.  Analytical theory for temporal damping of standing waves is developed by \cite{ruderman13}.  \cite{pascoe13} used analytical theory to do seismology on propagating MHD waves.  \cite{pascoe16} present a seismological study that uses values of the period, exponential damping time and Gaussian damping time. They make two statements that concern the essence of the seismological inversion problem.  First, that previous seismological studies were concerned with ill-posed problems while they deal with a well-posed problem.  Second, that their seismological study leads to a unique solution.  Both statements have to be interpreted with sufficient caution. If ill posed means that the problem has infinitely many solutions then the first statement is correct. The inversion problem of standing coronal loop oscillations has as a rule infinitely many solutions. As far as the second statement is concerned we note that  a seismological study that leads to a single solution has made additional assumptions on the variation of the equilibrium quantities and/or adopted values for specific characteristic quantities. But this single solution can not be unique.

\section{Limitations of coronal seismology and ill posed problems}

Let us have a look at the limitations of coronal seismology of standing MKD waves. The period $P$ and the exponential and Gaussian damping times, $\tau_{\rm D}$ and $\tau_{\rm G}$  are determined by the equilibrium model. In the $\beta =0$ approximation they are determined by the variation of the local Alfv\'en frequency. In case of a straight and constant magnetic field it is  the variation of density $\rho_0(r)$ and the strength of the constant magnetic field $B$ that determine the period $P$  and the damping times $\tau_{\rm D}, \tau_{\rm G}$ . Inspired by the success of helioseismology we might hope to determine the variation of $\rho_0(r)$ in the interval $[0, R]$. The  piece-wise constant density model with density $\rho_{\rm i}$ in the interior of the loop and density  $\rho_{\rm e}$ exterior to the loop was and still is popular in solar physics. Unfortunately it excludes resonant damping. It was replaced by a model in which  the density  $\rho_0(r)$ varies  in a continuous manner in a non-uniform layer   $R - l/2 \leq r \leq R + l/2$  with thickness $l$ from its interior value $\rho_{\rm i}$ to its exterior value $\rho_{\rm e}$.

The big difference between helioseismology and seismology of coronal loops is that in helioseismology many $(>10^6)$  modes are observed while in seismology of coronal loops there is one or occasionally two or more waves observed. The consequence is that  seismology of coronal loops is far from being able to determine the radial structure of equilibrium density  $\rho_0(r)$.   Only a  few characteristic quantities related to the equilibrium density can be determined.  It is out of the question to find a unique solution for the density distribution $\rho_0(r)$. The  classic seismological problem using observations and theoretical expressions for the periods and damping times of transverse standing MHD waves in coronal loops is better referred to as a reduced seismological problem.

The inversion of the reduced seismological problem has more unknowns than equations. The system is underdetermined and has infinitely many solutions. As we shall point out in what follows, there are four unknowns in what should be called a reduced seismological analysis. The unknowns are the longitudinal  Alfv\'en travel time $\tau_{\rm Ai}$, the density contrast $\zeta = \rho_{\rm i}/\rho_{\rm e}$, the radial inhomogeneity length scale $l$ and the steepness $\alpha$ of the density profile at the point of the resonance. This means that there are three equations for four unknowns and hence there is not a unique solution but $\infty^1$ solutions.  \cite{pascoe16} prescribe the value of $\alpha$.  They take $\alpha = 1$ for a linear variation. In the direct problem it is a good first attempt to adopt a linear variation for density to reduce the mathematical complexity in order to understand the physics. In the inverse problem the determination of the equilibrium density is a prime object of investigation and to assume from the start  that the variation of density is linear misses somehow the point of seismology. It also affects the inferred value of $l/R$. In reduced seismology it means that the number of unknowns is reduced from four to three. 


\section{The direct problem for period and exponential damping}
Let us consider the direct problem before focussing on the inverse problem. The direct problem is concerned with the analytical and/or numerical computation of periods and damping times of MHD waves of well defined theoretical equilibrium models. There is no discussion about the equilibrium distribution of the physical quantities and the geometric structure of the equilibrium. They are prescribed.  The equilibrium models that have played an important role in the study of MHD waves are the Cartesian slab model and the straight 1D cylindrical model.

In its initial version the Cartesian model has a piece wise constant density with density $\rho_{\rm i}$ to the left of $x=0$ and density $\rho_{\rm e}$  to the right of $x=0$. It has a constant magnetic field in the $z$-direction and constant plasma pressure. This equilibrium supports surface Alfv\'en waves. In the incompressible limit the square of the frequency of this surface Alfv\'en wave is the weighted mean of the squares of the local Alfv\'en frequencies

\begin{equation}
\sigma^2_{\rm SAW} = \frac{\displaystyle \rho_i \sigma^2_{\rm Ai} + \rho_{\rm e} \sigma^2_{\rm Ae}}{\displaystyle \rho_{\rm i} + \rho_{\rm e}},
\label{FreqSAW}
\end{equation}
where
\begin{equation}
\sigma^2_{\rm Ai,e} = \frac{\displaystyle B^2}{\displaystyle \mu \rho_{\rm i,e}}
\label{Freq1SAW}
\end{equation}
is the square of the local Alfv\'en frequency. The indices $\{{\rm i}, {\rm e}\}$ refer to $x<0$ and $x>0$ respectively. Expression 
(\ref{FreqSAW}) is given by for example \cite{chen74a,ionson78}; see also \cite{hasegawa82}. 
When the restriction of incompressibility is removed expression (\ref{FreqSAW}) is recovered for waves with $k_y >> k_z$. 

The initial version of the cylindrical model is a straight cylinder with piece wise constant density $\rho_{\rm i}$ for $0 \leq r \leq R$ and $\rho_{\rm e}$ for $r>R$ and constant magnetic field in the $z$-direction. The non-axisymmetric MHD waves with azimuthal wave number $m=1$ are of particular importance since they displace the axis and the magnetic tube as a whole. In the long wavelength or thin tube (TT) limit $k_z R << 1$ the frequency $\sigma^2_{KAW}$ of the fundamental radial mode of the non-axisymmetric waves is given by the same expression as that for the Cartesian surface Alfv\'en wave. 

The fact that the frequency of the fundamental radial mode of the non-axisymmetric waves with $m=1$ is equal to that of the surface Alfv\'en wave of the Cartesian piece wise constant density model is not surprising. \cite{wentzel79b} suggested that the non-axisymmetric waves of the cylindrical model were surface waves. \cite{goossens12a} gave a detailed discussion of the fundamental radial mode of the non-axisymmetric kink ($m=1$) waves and showed that they are indeed surface Alfv\'en waves with properties that are the same as those of the surface Alfv\'en waves of the Cartesian model. 
 
Expression (\ref{FreqSAW}) is given by equation (30) of \cite{spruit82} in the limit $k_z R  \rightarrow 0$. Expression (\ref{FreqSAW}) can be written as a simple expression for the period $P$  that depends on the density contrast $\zeta = \rho_{\rm i}/\rho_{\rm e}$ and the Alfv\'en travel time $\tau_{\rm Ai} = L/v_{\rm Ai}$ \citep[see][]{goossens08a}. Use the relations $\omega_k = k_z v_k, \;\;v_k = \sqrt{2} \;v_{\rm Ai}\; \rho_{\rm i} / (\rho_{\rm i} + \rho_{\rm e}), \;\; P = 2 \pi /\omega_k$ to obtain 
\begin{equation}
P = \tau_{\rm Ai} \;\sqrt{2} \;\left\{\frac{\zeta +1}{\zeta} \right \}^{1/2},
\label{Period1}
\end{equation}
with $L$ the length of the loop and $v_{\rm Ai}  = B /\sqrt{\mu \rho_{\rm i}}$.
Recall that the expression (\ref{Period1}) for the period  is for a model with a piece wise constant density. Expression (\ref{Period1}) implies that seismology based on observations of the period can only give information about $\tau_{\rm Ai}$ and $\zeta$.

The interest for the surface Alfv\'en wave was due to the possible heating because of the Alfv\'en resonance for thermonuclear fusion, application in geophysics and heating of solar and astrophysical plasmas. Early studies on resonant damping at the Alfv\'en resonance include \cite{chen74c,chen74b,chen74a,tataronis73,grossmann73,ionson78,hollweg88}. See a review in \cite{hasegawa82}.  However,  for resonant absorption to work the equilibrium has to be non-uniform.  Since 
\begin{equation}
\sigma^2_{\rm Ai} < \sigma^2_{\rm SAW} < \sigma^2_{\rm Ae}
\label{IneqFrSAW}
\end{equation}
the frequency of the surface Alfv\'en wave is in the Alfv\'en continuum,  the surface Alfv\'en wave undergoes resonant damping.  The non-uniformity was introduced into the equilibrium model by replacing the discontinuity at the surface $x=0$ in the Cartesian case or at the boundary surface $r=R$ in the cylindrical case by a continuous variation of equilibrium quantities in a transitional layer.  The local Alfv\'en frequency $\sigma_A$  varies in a continuous manner from its value $\sigma_{\rm Ai}$ to its value $\sigma_{\rm Ae}$ and defines the  Alfv\'en continuum. This can be done in several ways in the direct problem. \cite{chen74c,chen74b,chen74a,ionson78, hasegawa82} introduced a linear variation for the quantity $\epsilon$ defined as
\begin{equation}
 \epsilon(x) = \omega^2 \mu_0 \rho_{\rm 0}(x) - k_{\parallel}^2 B^2.
\label{epsilonx}
\end{equation}
The transitional layer is $[- a/2, \, a/2]$. 
When $B$ is constant Eq.~(\ref{epsilonx})  implies that the equilibrium density is a linear function in the transitional layer. \cite{hollweg88} take a constant magnetic field and adopt a linear variation for density in the transitional layer.  In the Cartesian case the plasma extends to the left of $x=0$ to $-\infty$ and to the right of $x=0$ to $+\infty$ . So there is no radius involved. \cite{hollweg88} in their numerical example  introduced $R$ by use of $k_y = m/R$ with $m=1$ for the non-axisymmetric kink wave. \cite{tataronis73} and \cite{grossmann73} adopted a linear variation for the equilibrium density $\rho_0(r)$ and for the square of the axial magnetic field $B_{\rm z,0}(r)$ so that the quantity $\epsilon(r)$ defined as 
\begin{equation}
 \epsilon(r) = \omega^2  \rho_{\rm 0}(r) - k_{\parallel}^2 B_{\rm z,0}^2 /\mu_0
\label{epsilonr}
\end{equation}
is a linear function of $r$. In these studies damping rates were obtained by adopting the so-called thin boundary approximation  \citep[see e.g.][]{hollweg88}.  The studies by  \cite{tataronis73} and \cite{grossmann73} were for incompressible motions. \cite{goossens92a} studied the damping of axisymmetric ($m=0$) (sausage) waves and non-axisymmetric waves. They adopted the thin tube (TT) approximation and the thin boundary (TB) approximation.
The use of the thin boundary (TB) approximation $l/R <<1, \lim l/R \rightarrow 0$ makes it possible to obtain
an analytical expression for $\tau_{\rm D}$. The big advantage of the thin boundary approximation is that numerical solutions of the ideal MHD equations are not required but analytical solutions can be used to the left and to the right of the resonant point. Connection formulae \citep[see e.g.][]{sakurai91,goossens95a,tirry96} are used to connect the ideal solutions across the resonant point for the determination of $\tau_{\rm D}$. The disadvantage of the TB approximation is that, strictly speaking,  it only works for $\lim\; l/R \rightarrow 0$. 

An important result of \cite{goossens92a} is their equation (77) which gives an  expression of the damping rate for damping due to resonant absorption for non-axisymmetric waves ($m \geq 1$). When corrected for a typo it is transformed in equation (30) of \cite{goossens09} 
\begin{equation}
\frac{\displaystyle \gamma}{\displaystyle \omega_{\rm k}} = - \frac{\displaystyle \pi/2}{\displaystyle \omega_{\rm k}^2} \frac{\displaystyle |m|}{\displaystyle R}\; \frac{\displaystyle \rho_{\rm i}^2 \;\rho_{\rm e}^2}{\displaystyle (\rho_{\rm i} + \rho_{\rm e})^3}\; \frac{\displaystyle (\omega_{\rm Ai}^2 - \omega_{\rm Ae}^2)^2}{\displaystyle \rho_{\rm 0}(r_A) \;\mid \Delta_{\rm A}(r_{\rm A}) \mid}.
\label{gamma1}
\end{equation}
In Eq.~(\ref{gamma1}) $\Delta_{\rm A}(r_{\rm A}) $ is 
\begin{equation}
\Delta_{\rm A} = \frac{d}{dr} (\omega^2  - \omega_{\rm A}^2) \mid_{r_{\rm A}}.
\label{DeltaA}
\end{equation}
In  the TB approximation the effect of the non-uniform layer on the damping is reduced to the value of $\Delta_{\rm A}$.
Equation~(\ref{gamma1}) was derived without making any assumptions about the variation of the equilibrium quantities in the non-uniform layer. Both density $\rho_{\rm 0}(r)$ and $B_{\rm z,0}(r)$ can vary with $r$. Consider an equilibrium with given values of $\rho_{\rm i}, \;\rho_{\rm e}, \; \omega_{\rm Ai}, \;  \omega_{\rm Ae}$. In the  thin tube (TT)  and thin boundary (TB)  approximations the damping rate due to resonant absorption is determined by the derivative of $\omega_{\rm A}^2(r)$ at $r_{\rm A}$  with $r_{\rm A}$ the position of the resonant point.  In the TB approximation it can be taken as $r_{\rm A} =R$.  Expression (\ref{gamma1}) indicates that observations of the damping rate can only give additional information on the value of $(d\omega_{\rm A}/dr)_{r_{\rm A}}$.

In what follows we assume that $B_{\rm z,0}$ is constant. This assumption  simplifies the mathematical expressions and it is a reasonable approximation of reality. For a constant axial magnetic field Equation~(\ref{gamma1}) can be simplified to (see equation [31] of \citealt{goossens09})
\begin{equation}
\frac{\displaystyle \gamma}{\displaystyle \omega_{\rm k}} = - \frac{\displaystyle \pi}{ \displaystyle 8}\; \frac{\displaystyle \mid m \mid }{\displaystyle R} \;\frac{\displaystyle (\rho_{\rm i} - \rho_{\rm e})^2}{\displaystyle (\rho_{\rm i} + \rho_{\rm e})}\;\frac{\displaystyle 1}
{\displaystyle \mid  (\frac{d \rho_{\rm 0}}{dr} )_{r_{\rm A}} \mid}.
\label{gamma2}
\end{equation}
Expression~(\ref{gamma2}) is remarkably simple.  In the case of a straight and constant magnetic field  the effect of the non-uniform layer is reduced to the value of the derivative of the equilibrium density at the resonant position
\begin{equation}
(\frac{d \rho_{\rm 0}}{dr})_{r_{\rm A}}. 
\label{DerivativeRho}
\end{equation}
The variation of density at the position $r_{\rm A}$ can be characterised by the dimensionless quantity $G$ defined as
\begin{equation}
G = \frac{\displaystyle R}{\displaystyle (\rho_{\rm i} - \rho_{\rm e})} \; \mid (\frac{d \rho_{\rm 0}}{dr})_{r_{\rm A}} \mid.
\label{G}
\end{equation}
$G$ relates the local variation of density $\rho_{\rm 0}$ at $r=r_{\rm A}$ to the global variation of $\rho_{\rm 0}$ from the value $\rho_{\rm i}$ to the value $\rho_{\rm e}$ over the distance $R$. The expression for the damping rate (\ref{gamma2}) (for $m=1$) can then be written as 
\begin{equation}
\frac{\displaystyle \gamma}{\displaystyle \omega_{\rm k}} = - \frac{\displaystyle \pi}{ \displaystyle 8}\; \frac{\displaystyle (\rho_{\rm i} - \rho_{\rm e})}{\displaystyle (\rho_{\rm i} + \rho_{\rm e})}\;\frac{\displaystyle 1}
{\displaystyle G}.
\label{gamma3}
\end{equation}
Now we can use the density contrast $\zeta = \rho_{\rm i} /\rho_{\rm e}$ and rewrite Eq.~(\ref{gamma3}) as an expression for the damping time $\tau_{\rm D}$ as 

\begin{equation}
\frac{\tau_D}{P} = \frac{4}{\pi^2}  \frac{\zeta +1}{\zeta -1} \;G.
\label{tauD1}
\end{equation}
Equation~(\ref{tauD1}) tells us that for a cylindrical straight equilibrium model with density contrast $\zeta$ the ratio of the damping time to the period is completely determined by the dimensionless quantity $G$. This is a remarkable result. All information on the variation of density is collapsed into this one quantity $G$. This is a big leap. From the point of coronal seismology it implies that the original seismological problem in the present formulation has infinitely many solutions. There is an unlimited number of density distributions that have the same value of the quantity $G$. 

In the direct problem the equilibrium configuration is freely chosen and the non-uniform variation of density can be confined to a layer of thickness $l$ with steepness $\alpha$. With that prescription 

\begin{equation}
\frac{d \rho_{\rm 0}}{dr} \mid_R = - \alpha  \frac{\rho_{\rm i} - \rho_{\rm e}}{l}.
\label{l-alpha}
\end{equation}
A linear variation of density corresponds to $\alpha =1$.  For a sinusoidal variation of density, introduced by \cite{ruderman02},  $ \alpha = \pi /2$. 

With the use of Eq.~(\ref{l-alpha})   expression  (\ref{G})  for $G$ becomes 
\begin{equation}
G = \frac{\displaystyle \alpha }{\displaystyle (l /R)}.
\label{G1}
\end{equation}
Equation~(\ref{G1}) implies  that there are infinitely many couples  $(\alpha, l/R)$ that produce the same value of $\tau_{\rm D}/P$. For example the value for $\tau_{\rm D}/P$ obtained for a linear variation for a layer with thickness $(l/R)_{L}$ is also recovered for the sinusoidal  profile and a layer with thickness $(l/R)_{L} \; \times \; \pi/2$. Conversely,  the value for $\tau_{\rm D}/P$ obtained for a sinusoidal variation for a layer with thickness $(l/R)_{S}$ is also recovered for the linear profile and a layer with thickness $(l/R)_{S} \; \times \; 2/\pi$. In the TTTB approximation different models can produce the same ratio  $\tau_{\rm D}/P.$  This does not affect the direct problem since  $\alpha,$ and $ l/R$  are prescribed. However, for the inverse problem this means that it is not possible to distinguish between different couples of $(\alpha, l/R)$ . 

With the use of expression~(\ref{G1}) we can rewrite Eq.~(\ref{tauD1}) as 
\begin{equation}
\frac{\tau_D}{P} = \frac{4}{\pi^2} \frac{\alpha}{l/R} \frac{\zeta +1}{\zeta -1}. 
\label{tauD2}
\end{equation}

Let us recapitulate what we have so far.  Equation~(\ref{Period1}) is an expression for the period of a piece wise constant density model in the long wave length or thin tube approximation. It depends on the density contrast $\zeta = \rho_{\rm i}/\rho_{\rm e}$ and the Alfv\'en travel time $\tau_{\rm Ai} = L/v_{\rm Ai}$ . It does not take into account a possible non-uniform variation of density. Equation~(\ref{tauD1}) has been derived in the thin tube and the thin boundary (TTTB) approximations. It shows that the damping time for exponential damping due to resonant absorption depends on $\zeta$ and $G$, with  $G$ the ratio of $\alpha$ and $l/R$.

In the direct problem the approximations of thin tube and thin boundary can be dropped and numerical solutions can be obtained for the periods, the damping times and eigenfunctions of the resonantly damped MHD waves. For general equilibrium models  it is in general no longer possible to obtain analytical solutions, but in the direct problem that does not matter.

Expression~(\ref{Period1}) for the period neglects the effect  of the variation of the density in the non-uniform layer. This effect is not negligible a priori, in particular when the layer is thick. The effect of the non-uniform layer on the period has been studied by \cite{soler13}.

A common argument to use the thin boundary approximation for the damping time is that the numerical simulations by \cite{vandoorsselaere04a} and \cite{arregui05} show that the dependence of $\tau_{\rm D}$ on the width of the non-uniform layer is not very strong. \cite{arregui08a} have shown, in the context of prominence threads oscillations, that different combinations of wavelength, density contrast and width of the non-uniform layer lead to percentage deviations up to 20\% with respect to the thin boundary approximation for the damping. However, the numerical simulations by \cite{vandoorsselaere04a}, \cite{arregui05} and \cite{arregui08a} were for a sinusoidal variation of density.  \cite{soler13} have shown that  other variations of density result in a different  dependence on the width.  \cite{soler14a} and \cite{arregui15c} discussed the consequences of different density profiles on coronal seismology for a scheme using period and exponential damping times.

\section{The direct problem for Gaussian damping}

Gaussian damping is first studied for spatial damping of propagating waves by \cite{pascoe12} in numerical simulations. Analytical theory for spatial damping of  propagating waves is developed by \cite{hood13}.  They  studied the resonant damping of propagating kink waves in a semi-infinite magnetic tube, where the waves are excited by a driver at the tube end. They found that the wave amplitude is approximately described by a Gaussian function at sufficiently small distances from the tube end.  At larger distances they found that the wave amplitude decays exponentially. Also  the distance where the transition from the Gaussian to the exponential damping occurs,  increases with the width of the non-uniform layer.  Finally, they found that the classical theory of resonant absorption overestimates the damping distance and suggests a slower damping.  An expression for the Gaussian damping distance  $L_{\rm G}$ for small $k_{\rm z}z$ is given in their equation (49).  We note that \cite{hood13} use the TTTB approximation and also a linear variation of density in the non-uniform layer ($\alpha = 1$).

Analytical theory for the temporal damping of standing waves is developed by \cite{ruderman13}. Similarly to \cite{hood13}, \cite{ruderman13} use the TTTB approximation and  a linear variation of density in the non-uniform layer ($\alpha = 1$).  An analytical expression for the Gaussian damping time $\tau_{\rm G}$ is not given by \cite{ruderman13}.  They solve their governing equation (30) and compare numerical values of the damping time  with the damping time predicted by the classical theory of resonant absorption, $\tau_{\rm D}$. They find that $\tau_{\rm D}$  exceeds the corresponding numerical values of the actual damping time by about 9\% for $\epsilon  = 0.1$, about 14\% for $\epsilon = 0.2$, and about 18\% for $ \epsilon  = 0.3$ . The quantity $\epsilon$ here is different from what is used in Eqs.~ (\ref{epsilonx})  and   (\ref{epsilonr}).  Here it is the ratio of the thickness of the non-uniform layer $l$ to the radius $R$:    $\epsilon = l/R$.
The  classical theory of resonant absorption underestimates the actual damping as the classical damping time is longer than the corresponding Gaussian damping time.  The error increases  when $\epsilon$  increases. However, in the view of \cite{ruderman13}  the error is not very large, and  the damping time given by the classical theory of resonant absorption can be taken as a reasonable approximation.

\cite{ruderman13} note that there is an obvious similarity between their results and those obtained by \cite{hood13}  when distance is interchanged with time.  The oscillation amplitude is approximately described by a Gaussian function for sufficiently short time.  For standing waves the oscillation amplitude decays exponentially at later times and the time when the transition from the Gaussian to the exponential amplitude profile occurs increases for larger values of $\epsilon$.  Finally,  the classical theory of resonant absorption underestimates the actual damping.

Analytical studies of Gaussian damping have used the TTTB approximation and also a linear variation of density in the non-uniform layer ($\alpha = 1$).  There are no analytical expressions for the Gaussian damping time $\tau_{\rm G}$ or interpolation formulae for that quantity derived from results of numerical simulations.

\section{Seismology for exponentially damped standing MHD waves}

Let us have a look at seismology with classical exponential damping.  First,  consider observed values  of period and the corresponding Eq.~(\ref{Period1}).  There is one equation and two unknowns: the density contrast  $\zeta$ and the Alfv\'en travel time $\tau_{\rm Ai}$. Hence there are $\infty^1$ solutions. The problem is ill-posed since there are more unknowns (two) than observed quantities (one).  When the  density contrast $\zeta$ is prescribed  one solution for $\tau_{\rm Ai}$ is found. This is done by \cite{nakariakov01}.   They took  $\zeta  = 10$ and found one solution for $\tau_{\rm Ai}$. \cite{nakariakov01} choose a specific value for $L$ and for $\rho_{\rm i}$ to determine an estimate for $B$.

Secondly, consider observed values of $\tau_{\rm D}$ and use Eq.~(\ref{tauD1}). There is one equation. The unknowns are 
the density contrast $\zeta$ and the quantity $G$, with $ G = \alpha/(l/R)$, so that there are three unknowns:  the density contrast $\zeta$, 
the radial inhomogeneity length scale,  $l/R$,  and the steepness of the density variation $\alpha$.  Hence the reduced seismological inversion scheme has  $\infty^2$ solutions. The problem is ill-posed because there are more unknowns (three) than observed quantities (one).  When the  density contrast $\zeta$ and $\alpha$  are prescribed one solution for $l/R$ is found.  This is done by \cite{ruderman02} and \cite{goossens02a}. Both investigations  used  $\zeta = 10$ and assumed that the density variation in the intermediate non-uniform layer is sinusoidal, so that  $\alpha = \pi/2$. \cite{ruderman02} considered one numerical example to proof the principle. \cite{goossens02a}  computed $l/R$ for 11 events. The same results as far as the damping times are concerned  can be obtained with the use of a linear variation of density and non-uniform layers that are thinner by a factor $2/\pi$ than those with the sinusoidal variation. 

Thirdly, consider observed values of $P$ and  $\tau_{\rm D}$ and use Eqs.~(\ref{Period1}) and  (\ref{tauD1}).  There are two equations. 
The unknowns are the Alfv\'en travel time $\tau_{\rm Ai}$,  the density contrast $\zeta$ and the quantity $G$, with $ G = \alpha/(l/R)$, so that there are  four unknowns:  the Alfv\'en travel time $\tau_{\rm Ai}$,  the radial inhomogeneity length scale,  $l/R$,  the density contrast $\zeta$ and the steepness of the density variation $\alpha$.  Hence there are  $\infty^2$ solutions. The problem is ill-posed since there are more unknowns (four)  than observed quantities (two). \cite{arregui08a} (fully numerical) and \cite{goossens08a}  (analytical) removed one unknown, namely $\alpha$,  from the analysis by adopting a fixed sinusoidal variation of density so that  $\alpha = \pi/2$. That left them with two equations for three unknowns and hence there are $\infty^1$  solutions. We note that \cite{arregui07a} solved the dissipative MHD equations without making the TT approximation and the TB approximation.   Again in the context of the TTTB approximations the results of \cite{arregui07a} and \cite{goossens08a} can be recovered by the use of a linear variation of density and non-uniform layers that are thinner by a factor $2/\pi$ than those with the sinusoidal variation. Actually, other prescriptions of the variation of density; meaning that other values of $\alpha$  can be used together with the corresponding values of $l/R$. 

\section{Seismology for Gaussian damped standing MHD waves}

\cite{pascoe13} used analytical theory to do seismology of propagating waves.  Their Figure 2 shows that Gaussian damping is only important for low density contrasts and for a short distance. For a density contrast of $\zeta=3$, Gaussian damping is present for only two (or fewer) wavelengths. The statement that the exponential regime is present for long /distances is too strong. It is already there after two (or fewer) wavelengths.  As far as Gaussian damping in time is concerned, the last two paragraphs of Section 4 and Section 5 of \cite{ruderman13} are very helpful. They point out the similarities between spatial and temporal damping.  In particular ``the oscillation amplitude is approximately described by the Gaussian function at sufficiently small time''.``The damping time given by the classical theory can be used as a reasonable approximation''.

\cite{pascoe16} do seismology of standing MHD waves. They introduce a third quantity, the Gaussian damping time $\tau_{\rm G}$, and claim that the inversion problem is now well posed. This claim is not correct. First, they forget that they are dealing with a reduced seismological problem and that the full seismological problem a fortiori has not a unique solution. Secondly, not even their reduced seismological problem has a unique solution.  There are three observed quantities: the period $P$  and the damping times $\tau_{\rm D}, \tau_{\rm G}$ and four unknowns: the Alfv\'en travel time $\tau_{\rm Ai}$, the density contrast $\zeta$, the radial inhomogeneity length scale $l/R$ and the steepness of the density profile at the point of the resonance $\alpha$.   Hence there are $\infty^1$ solutions.  In the same manner as \cite{arregui07a} and \cite{goossens08a},  \cite{pascoe16} prescribe the value of $\alpha$.  They take $\alpha = 1$ for a linear variation. Since $\alpha$ is given a prescribed value, a single solution for the three remaining unknowns is found. A different choice for $\alpha$ would lead to a different solution for the three remaining unknowns. 

For the period $P$ and the exponential damping time $\tau_{\rm D}$ \cite{pascoe16} use the classic expressions obtained in the TTTB approximation for a linear profile  ($\alpha = 1$).  In order to obtain an expression  for $\tau_{\rm G}$ \cite{pascoe16} use the relation (see their equations [4] and [5])

\begin{equation}
\frac{\tau_{\rm G}}{P} = \frac{L_{\rm G}}{\lambda},
\label{TauGLg1}
\end{equation}
with $L_{\rm G}$ the distance of Gaussian damping of propagating waves and $\lambda$ the wavelength.  For exponential damping the damping time $\tau_{\rm D}$ for standing waves can be related to the damping length $L_{\rm D}$ of propagating waves by 
\begin{equation}
\frac{L_{\rm D}}{\lambda} = \frac{v_{\rm gr}}{v_{\rm ph}}  \frac{\tau_{\rm D}}{P},
\label{TauDLD}
\end{equation}
with  $v_{\rm gr}= \partial \omega / \partial k_{\rm z}$  the group velocity and $v_{\rm ph} = \omega / k_{\rm z}$ the phase velocity.  When the frequency $\omega$ varies linearly with $k_{\rm z}$ then $v_{\rm gr}/ v_{\rm ph} = 1$.  It is not clear whether this relation (\ref{TauDLD}) also holds for Gaussian damping.  Nowhere could we find  any evidence in that respect in the literature.  So far we have been unable to derive an analytical expression for $\tau_{\rm G}$.

\section{Seismology from Bayesian inference}

So far our discussion was focused on the mathematical procedure to obtain algebraic estimates for unknown parameters from observable wave properties. This section provides arguments from the perspective of Bayesian inference. Bayesian inference is increasingly applied to astrophysical problems \citep{asensioramos18} and is also implemented in the context of coronal seismology \citep[see review by][]{arregui18}.

When confronted with the solution of an inverse problem \citep[see][for an in-deep discussion on inverse problems]{tarantola87}, the first difficulty is to find at least one model (in our case, a discrete set of parameter values) that is consistent with the data (a discrete set of observables). The discussion above has shown that this is not possible when the amount of unknown parameters outnumbers that of observables. The situation will of course depend on what exactly one means by consistent with the data and by the amount of information one is willing to accept (as certain) about the particular value of a given model parameter. Recall that \cite{pascoe16}, in the same manner as \cite{arregui08a} and \cite{goossens08a}, prescribe the value of $\alpha$, by adopting a particular profile for the variation of density in the non-uniform layer. A second difficulty, by all means more complicated that the first,  is to quantify the uniqueness of the obtained solution. This is the reason why most applied inverse problems across different research areas use optimisation methods trying to find a so-called best fit model.

The Bayesian approach to the solution of inverse problems gets rid of these issues by assuming from the outset that there is no such a thing as a best fit model. By accepting that because the inversion procedure must use information that is incomplete and uncertain the best we can do is to quantify the degree of plausibility of alternative parameter values/models.  This degree of plausibility can only be relative. Hence the solution to the inverse problem cannot be unique and is expressed by probability density functions that tell us how the plausibility is distributed over the considered parameter values/models.

This being said, let us now discuss how the Bayesian framework helps us support our mathematical arguments while at the same time offers some additional insight into the relationship between the algebraic and probabilistic results. We do so by computing posterior probability density distributions, $p$({\boldmath $\theta$}$| M,D$), for the vector of unknown parameters {\boldmath$\theta$}, conditional on the assumed theoretical model $M$ and the observed data $D$. Once the full posterior is computed, marginal posteriors for each unknown in the parameter vector can be computed by marginalisation.

\begin{figure*}[t]
\center
\includegraphics[width = 0.49\textwidth]{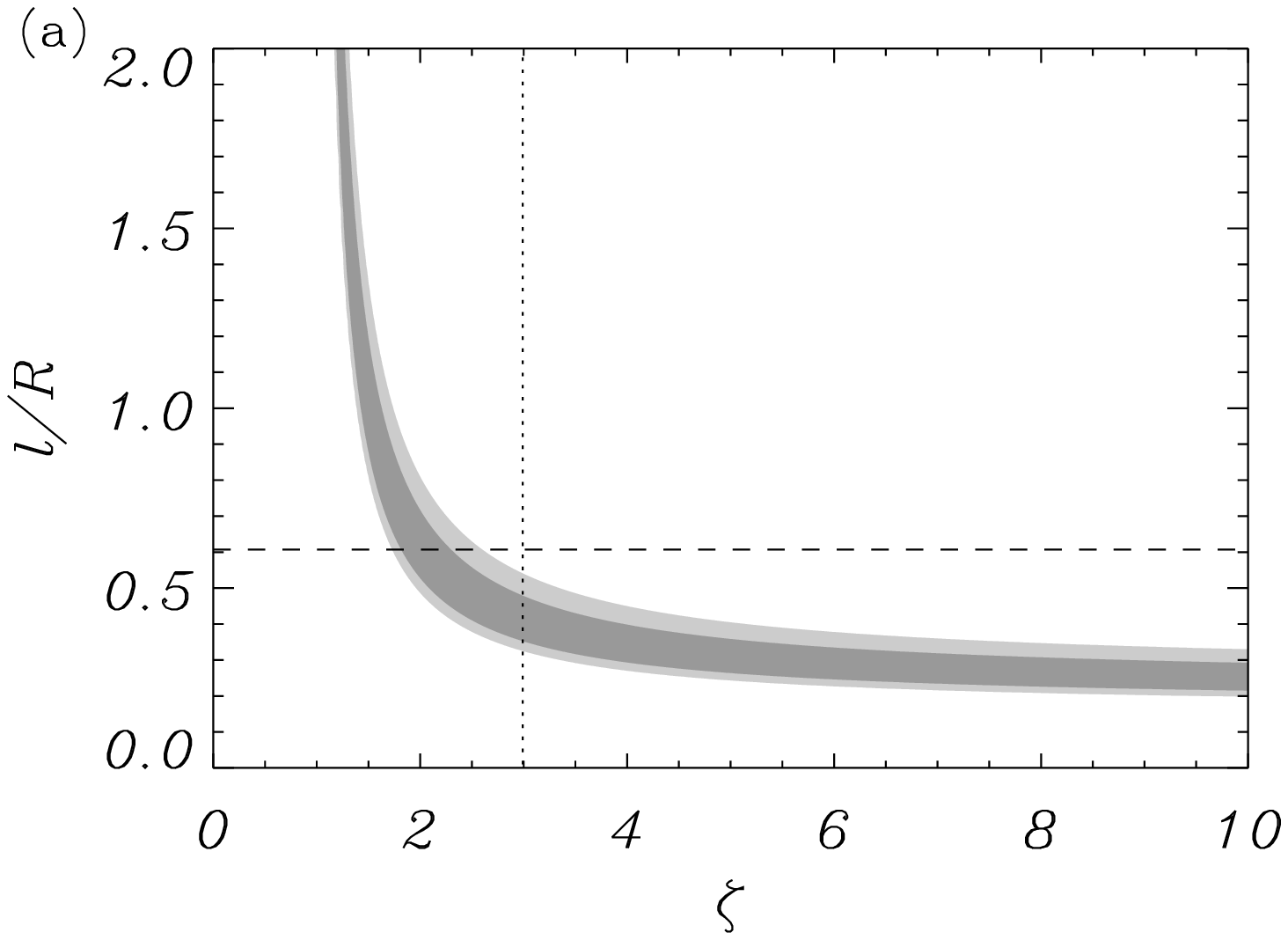} 
\includegraphics[width = 0.49\textwidth]{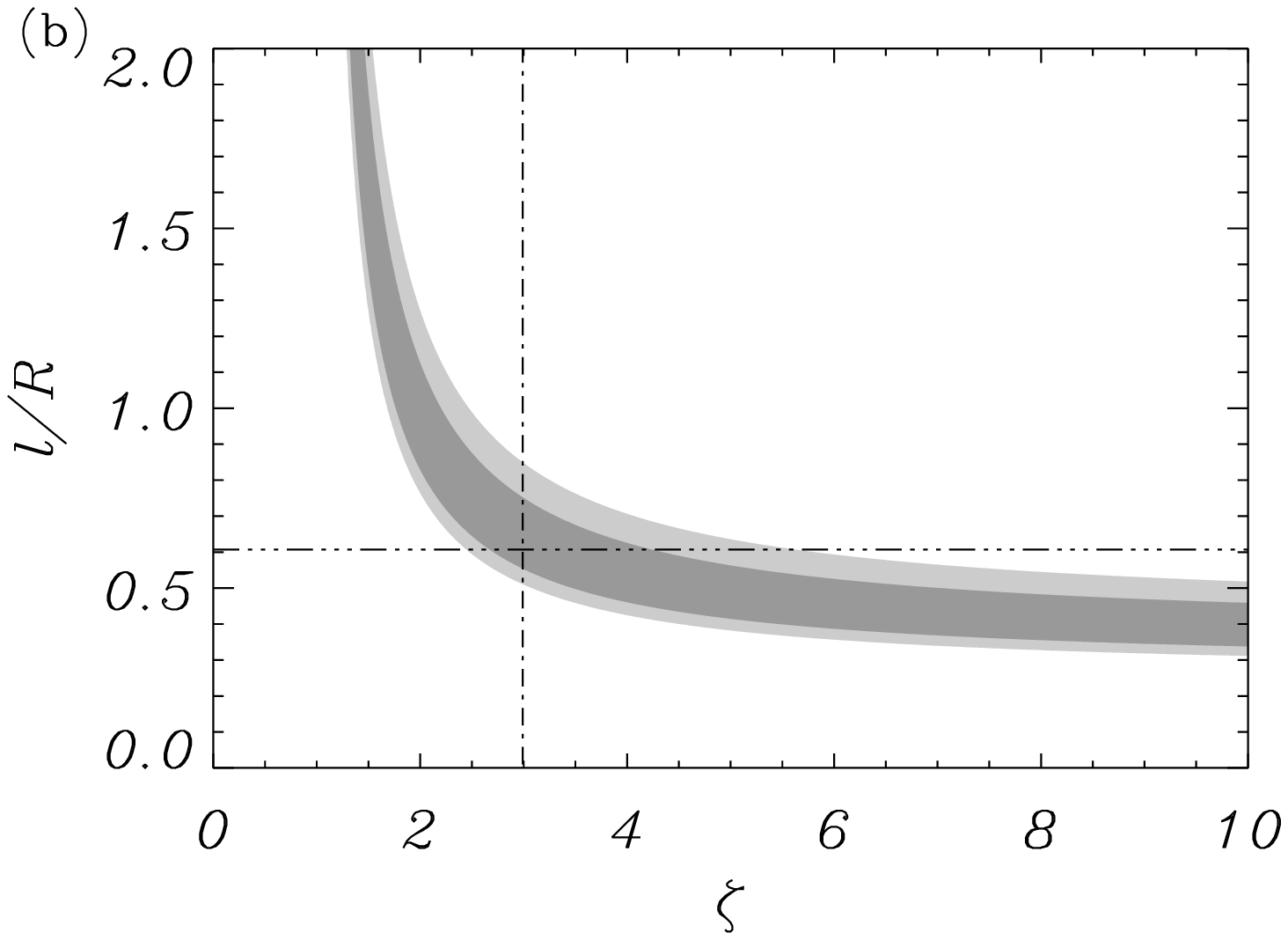} \\
\includegraphics[width = 0.49\textwidth]{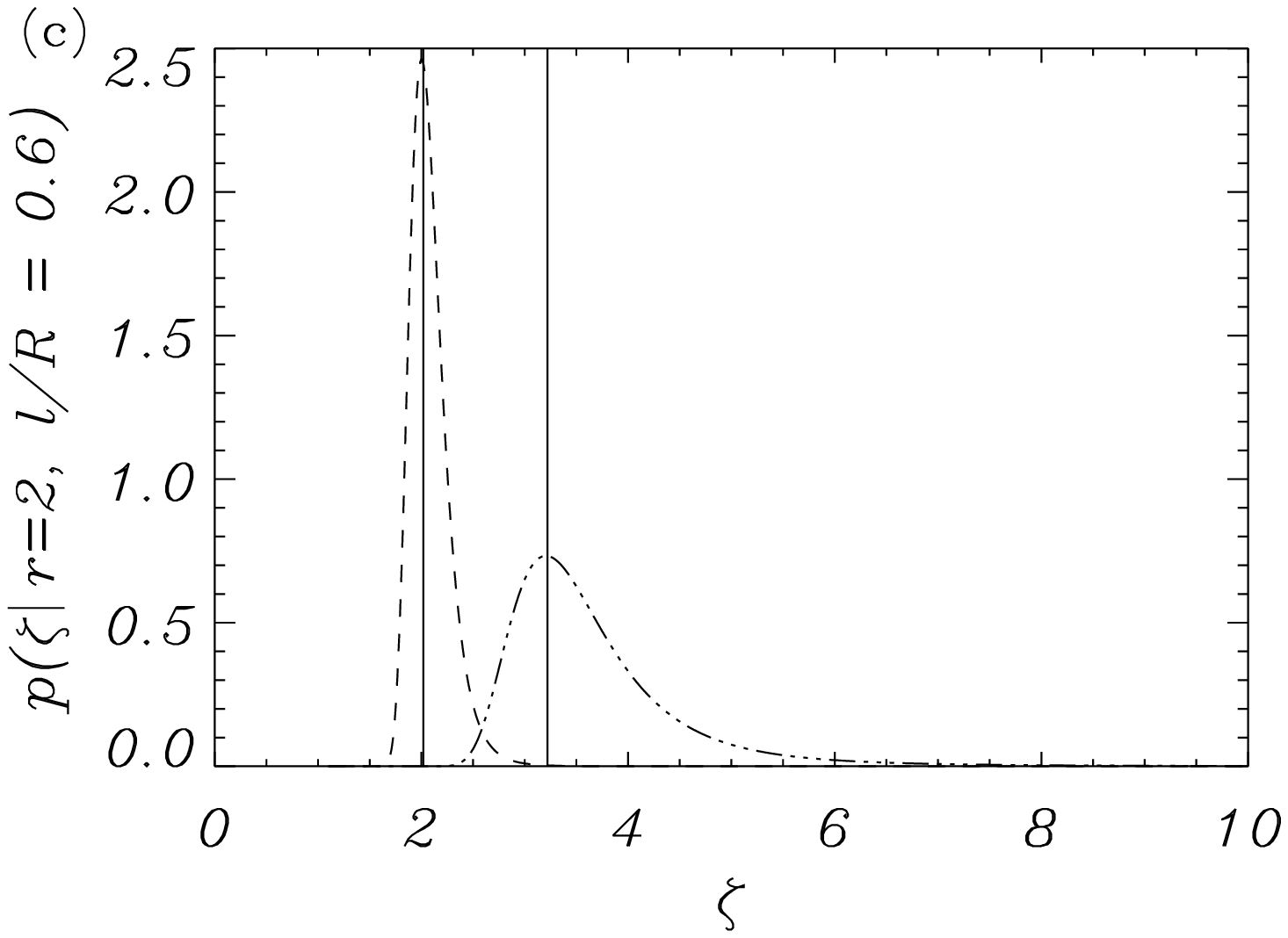} 
\includegraphics[width = 0.49\textwidth]{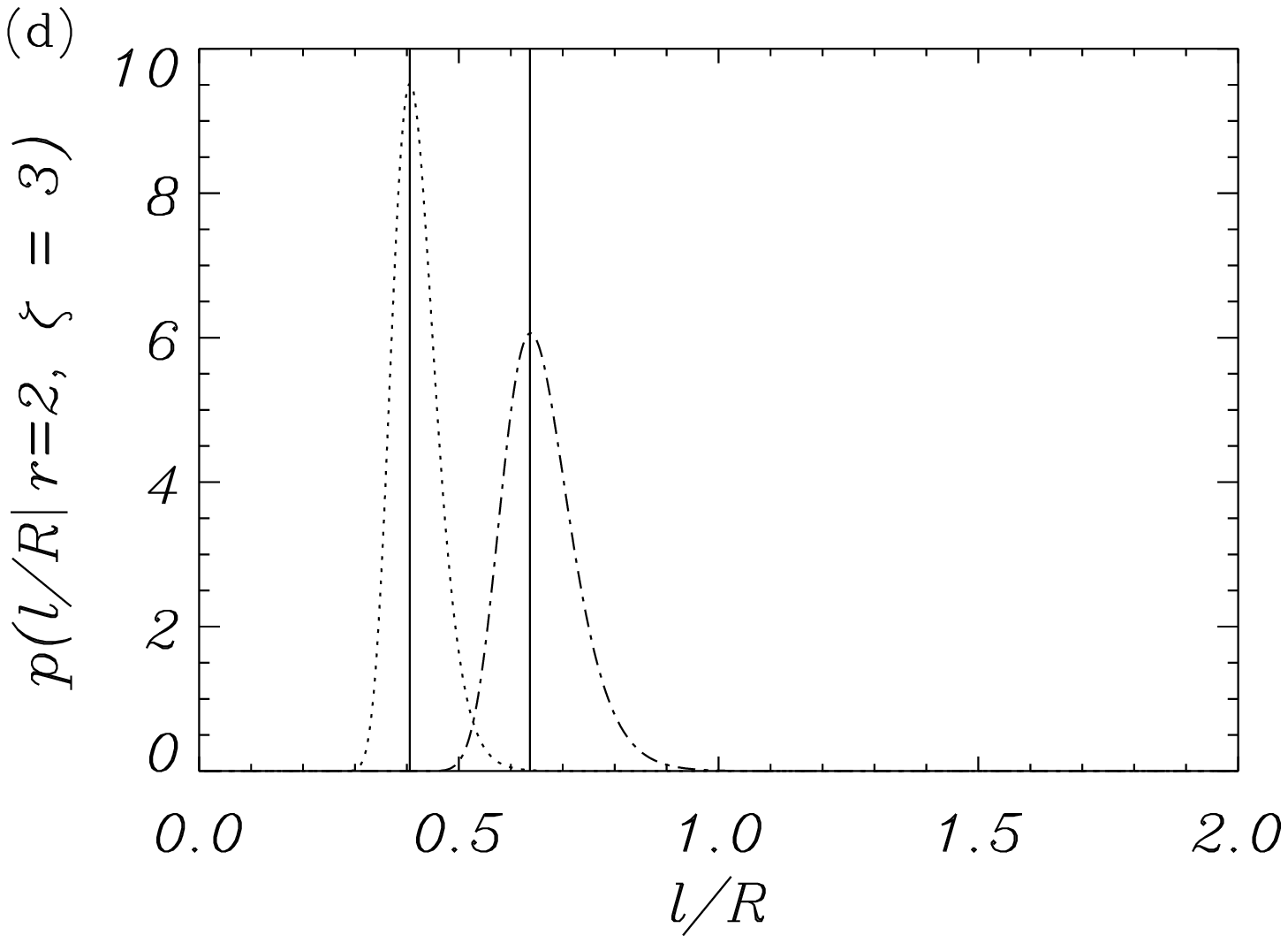}\\
\includegraphics[width = 0.49\textwidth]{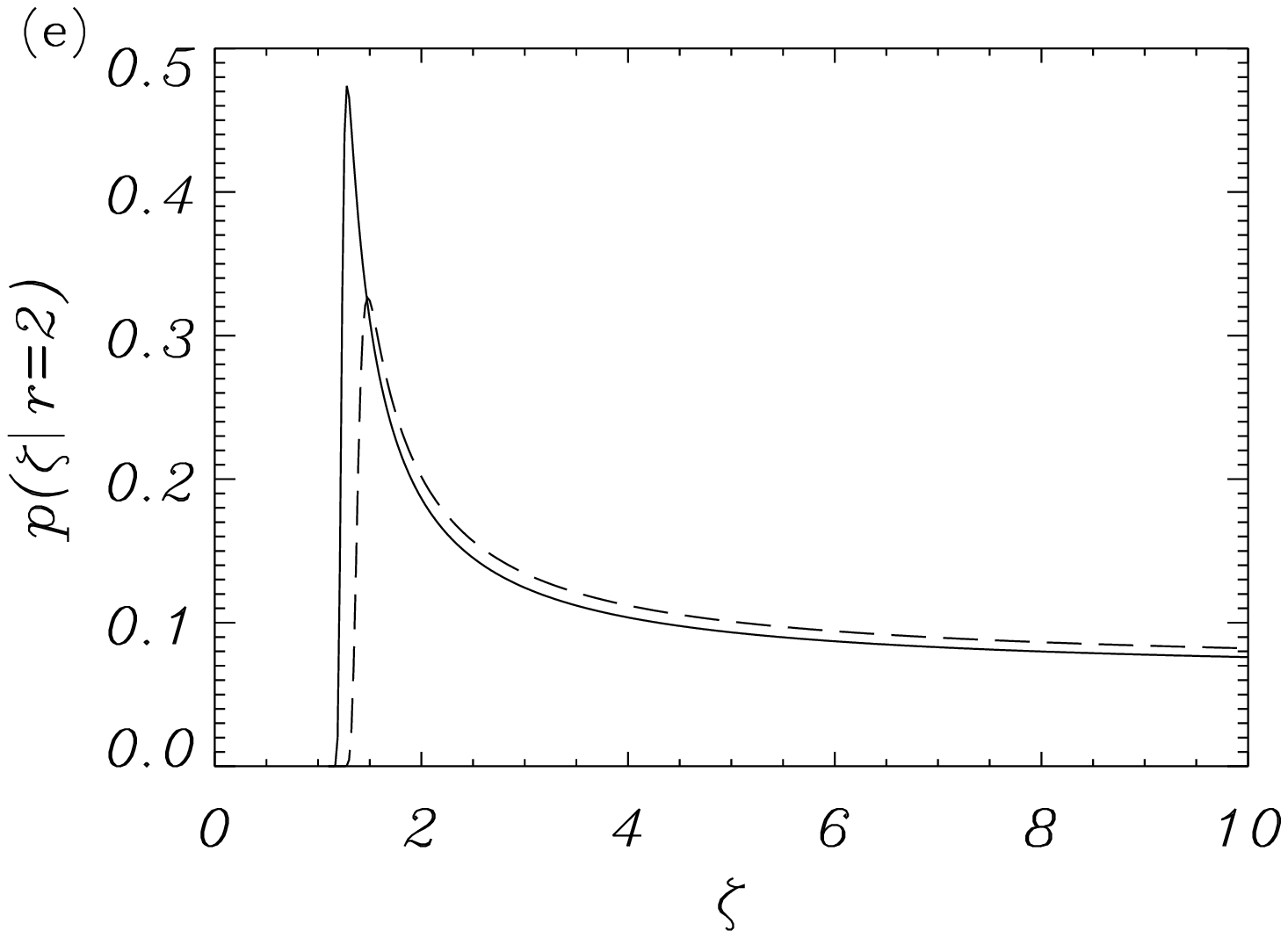} 
\includegraphics[width = 0.49\textwidth]{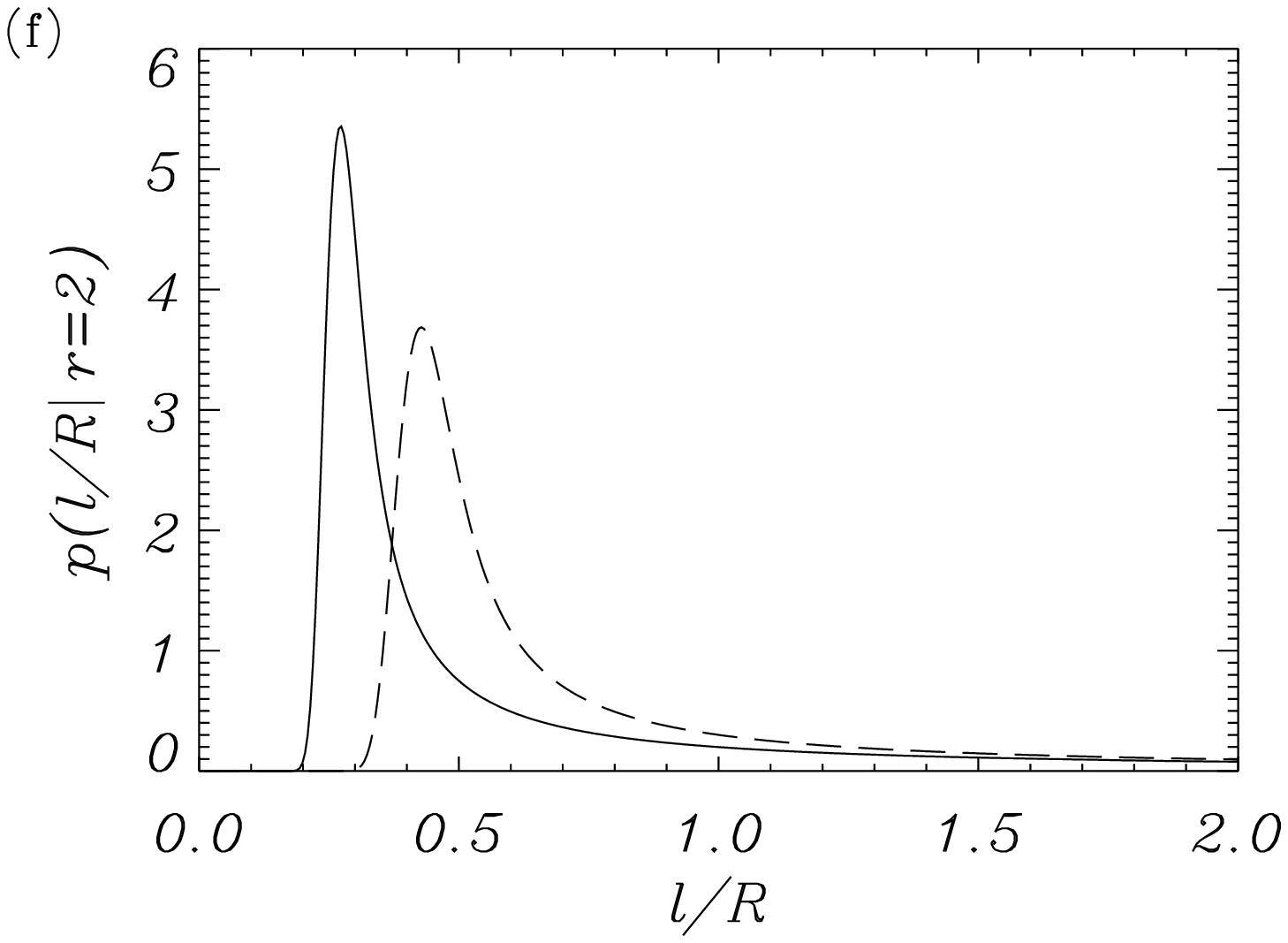}
\caption{Results from the comparison of inferences using two values of $\alpha$ in Eq.~(\ref{tauD2}). (a) and (b) show the two-dimensional joint posteriors in the ($\zeta$, $l/R$) parameter space for $\alpha=1$ and $\alpha=\pi/2$, respectively. (c) and (d) show one-dimensional cuts of the joint posteriors along fixed values of $l/R=0.6$ and $\zeta=3$, respectively. Same line-styles are used to identify cut directions in (a) and (b) with the corresponding results in (c) and (d). The vertical solid lines in (c) and (d) show the algebraic inversion results using Eq.~(\ref{tauD2}) and the fixed values of $l/R=0.6$ in panel (c) and $\zeta=3$ in panel (d). (e) and (f) show the marginal posteriors for $\zeta$ and $l/R$, respectively. In these calculations $r=\tau_{\rm D}/ P=2$ and $\sigma=0.1r$.\label{bayestwoalphas}}
\end{figure*}

\subsection{Seismology from exponential time damping}

The Bayesian solution to the problem of inferring the two parameters that define the cross-field density structure ($\zeta$ and $l/R$) from the damping ratio of resonantly damped oscillations is presented by \cite{arregui11b} (by numerical sampling of the posteriors using Markov Chain Monte Carlo methods) and \cite{arregui14} (by direct numerical integration). In both studies, a value of $\alpha=\pi/2$ (sinusoidal density profile) is adopted. \cite{arregui15c} showed how, in strong damping regimes, inference results computed with different profiles for the density in the non-uniform layer can substantially differ. They also showed how obtaining evidence in favour of a particular density model using Bayesian model comparison can be difficult from the practical point of view. The reason is that sufficiently large Bayes factors in support of a given model can only obtained in very strong damping regimes.

We consider the Bayesian inversion of the direct problem given by Eq.~(\ref{tauD2}) for two density models,  $\alpha =1$ (linear density profile) and $\alpha=\pi/2$ (sinusoidal model), by taking a hypothetical observed damping ratio with its measurement error, $D=\{\tau_{\rm D}/P\}= 2.0\pm0.2$, uniform priors for the two unknowns, {\boldmath$\theta$}=$\{\zeta, l/R\}$, over the ranges $\zeta\in[1.1,10] $ and $l/R\in[0.01,2]$, and a Gaussian likelihood function \citep[as described in][]{arregui14}. The top two panels in Fig.~\ref{bayestwoalphas} display the joint probability density function for $\zeta$ and $l/R$ for the case $\alpha = 1$ (Fig.~\ref{bayestwoalphas}a) and $\alpha = \pi/2$ (Fig.~\ref{bayestwoalphas}b). This joint probability density, $p$($\{\zeta,l/R\}$| $M,D$), expresses how the plausibility of different combinations of $\zeta$ and $l/R$ is distributed in the two-dimensional parameter space of unknowns. The dark and light grey shaded regions indicate the 68\% and 95\% credible regions, respectively. They delimit the boundaries of the areas with combinations of $\zeta$ and $l/R$ with probabilities larger that those two particular values. It is clear that, although similar, these two plausibility distributions for different values of $\alpha$ are different, that is $p$($\{\zeta,l/R\}$| $M_{\rm \alpha=1},D$) $\neq$ $p$($\{\zeta,l/R\}$| $M_{\rm \alpha=\pi/2},D$). Even if we consider separately each panel (a or b), corresponding to a given $\alpha$ (1 or $\pi/2$), the solution is not unique. Some combinations are simply more plausible than others. 

\begin{figure*}[t]
\center
\includegraphics[width = 0.49\textwidth]{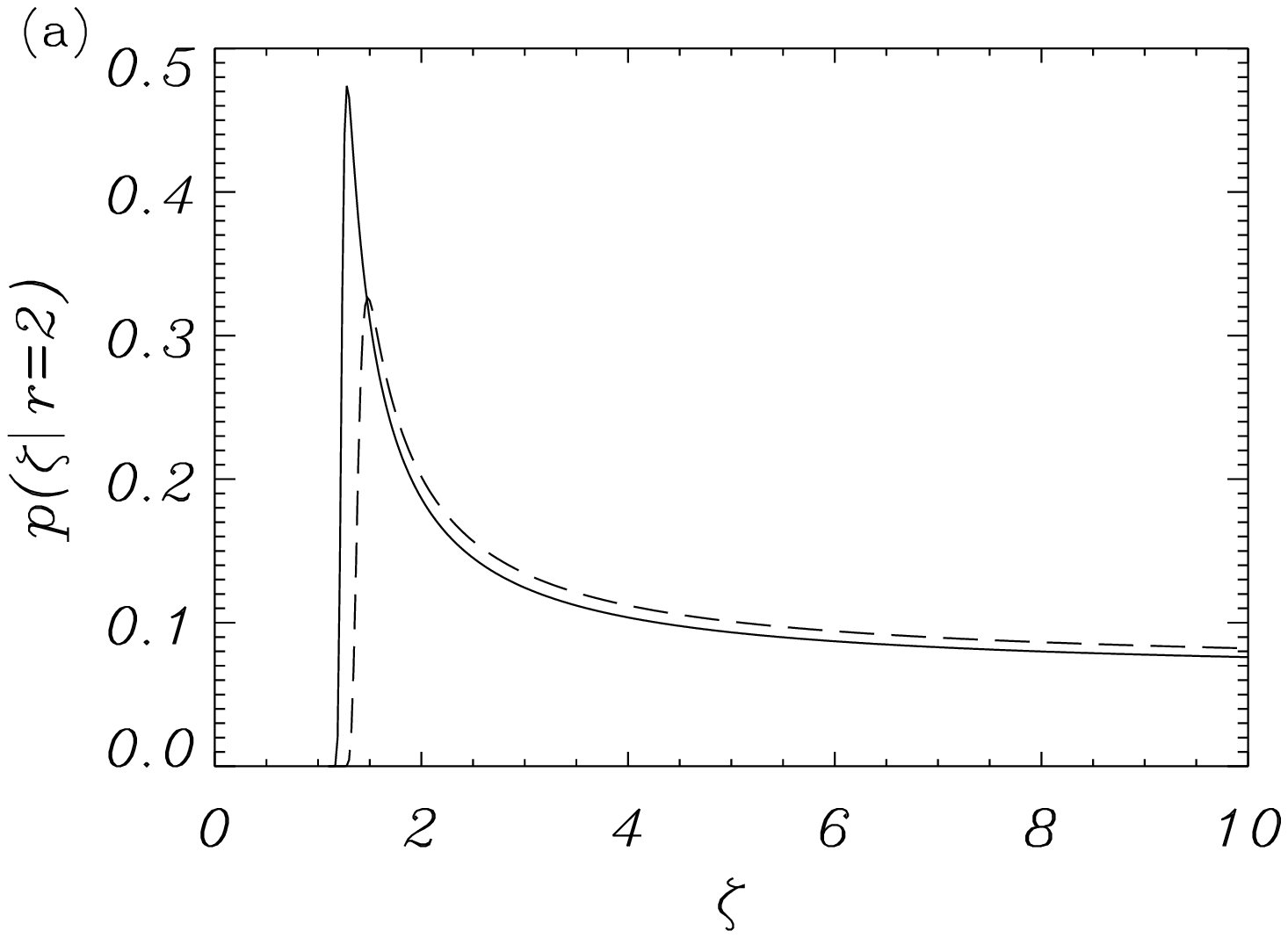} 
\includegraphics[width = 0.49\textwidth]{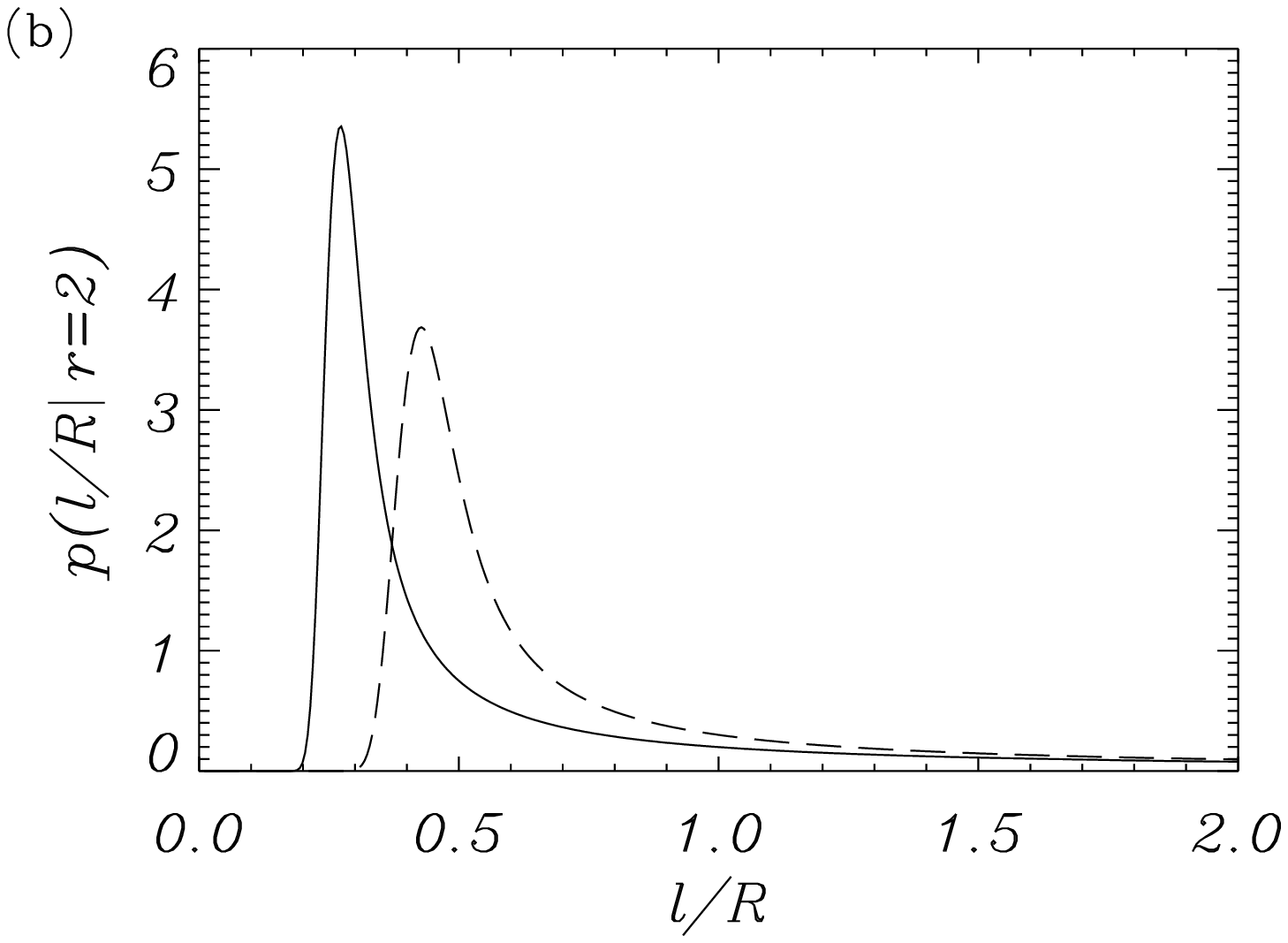} \\
\includegraphics[width = 0.49\textwidth]{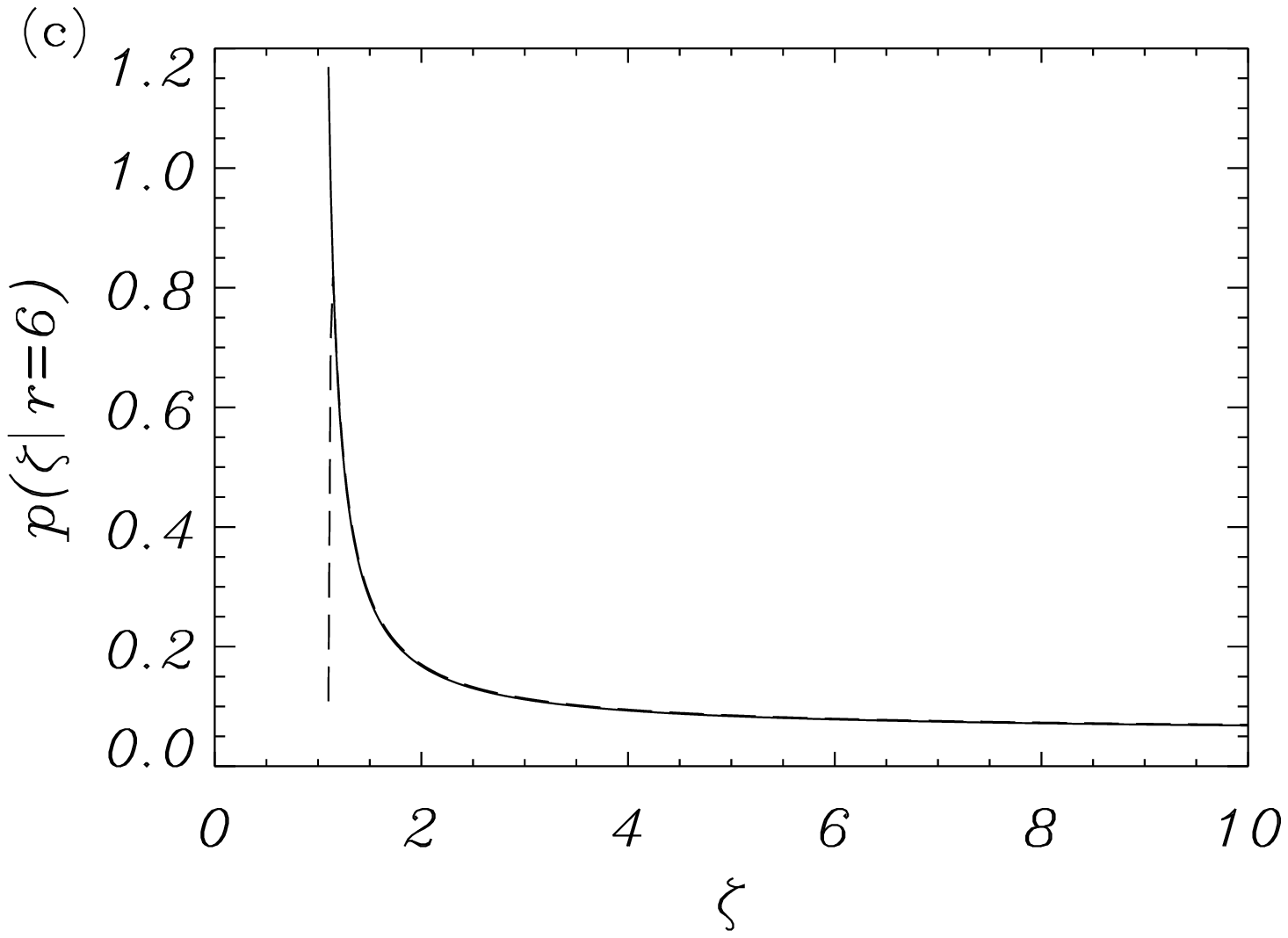} 
\includegraphics[width = 0.49\textwidth]{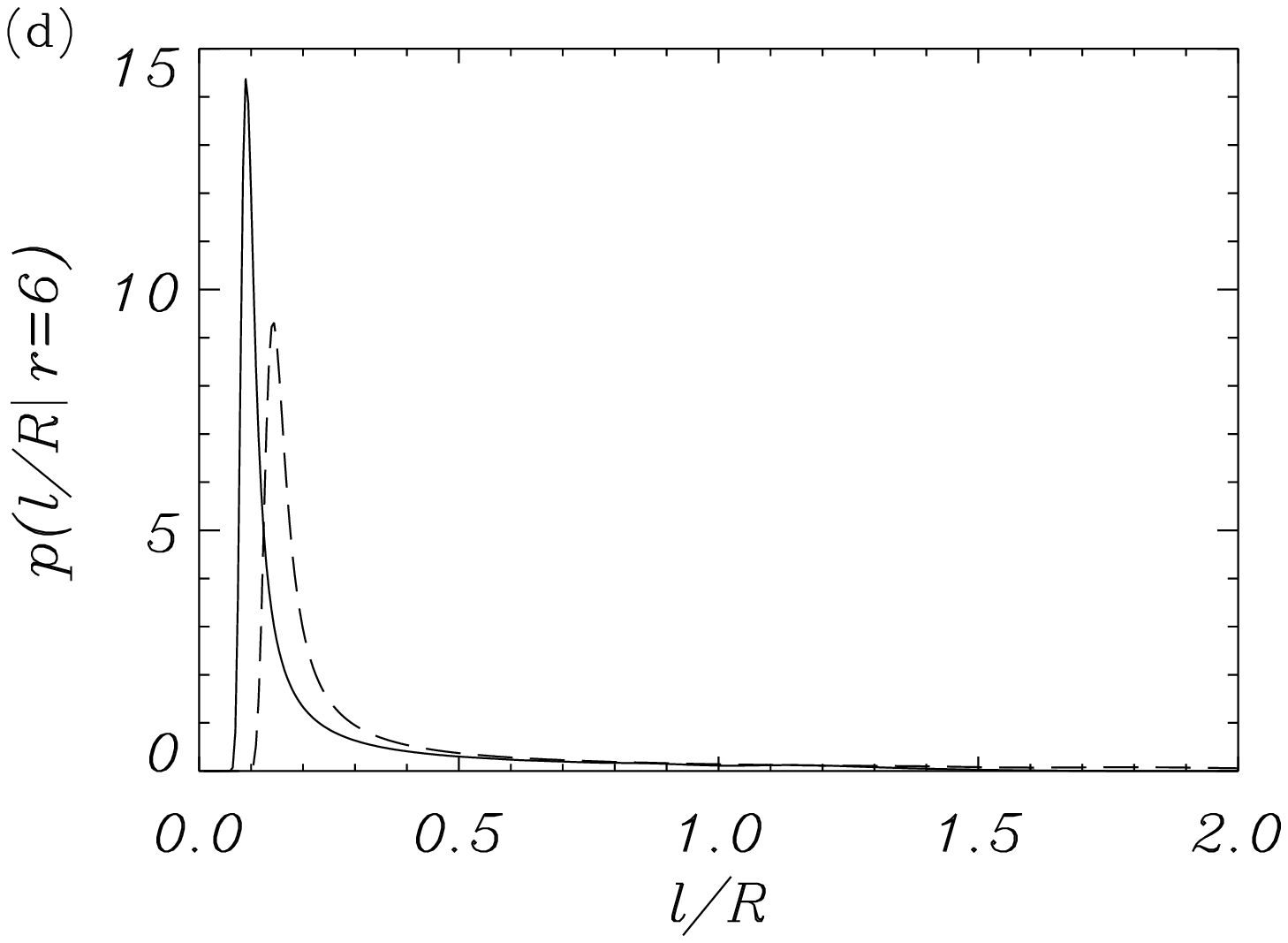}
\caption{\label{fig2} Marginal posteriors for $\zeta$ and $l/R$ from the comparison of inferences using two values of $\alpha$ and for two values of damping ratio: (a) and (b): 
$r=\tau_{\rm D}/ P=2$; (c) and (d): $r=\tau_{\rm D}/ P=6$. In both cases $\sigma=0.1r$.\label{diffwithdampingrate}}
\end{figure*}

In algebraic seismology using discrete values for observables and parameters it has been common practice to adopt a fixed value for one of the parameters, thus reducing the number of unknowns. For example, one could fix the value of $\zeta$ in Eq.~(\ref{tauD2}) and solve for $l/R$. We can mimic this procedure in our probabilistic inference example by taking cuts of the joint probability density functions displayed in Figs.~\ref{bayestwoalphas}a and b along fixed values of $\zeta$ or $l/R$. By so doing, we obtain the one-dimensional posterior density functions displayed in the two middle panels of Fig.~\ref {bayestwoalphas}.  Figure~\ref{bayestwoalphas}c shows posteriors along cuts at a fixed value of $l/R$, $p$($\zeta$| $l/R=0.6,M_{\rm \alpha=1},D$) and $p$($\zeta$| $l/R=0.6,M_{\rm \alpha=\pi/2},D$). Fig.~\ref{bayestwoalphas}d shows posteriors along cuts at fixed value of $\zeta$, $p$($l/R$| $\zeta=3, M_{\rm \alpha=1},D$) and $p$($l/R$| $\zeta=3, M_{\rm \alpha=\pi/2},D$). 

Let us first focus on Fig.~\ref{bayestwoalphas}c. By assuming a given value for $l/R$, two one-dimensional posteriors for $\zeta$ are found, one for each value of $\alpha$. It is clear that they differ, both in central tendency and dispersion. There is shift in the maximum a posteriori estimates. Even if we were to consider only one of the two posteriors in Fig.~\ref{bayestwoalphas}c, the solution is not unique. Some values of $\zeta$ are simply more plausible than others. 

We can repeat the same arguments based on the results displayed in Fig.~\ref{bayestwoalphas}d. By assuming a given value for $\zeta$, two one-dimensional posteriors for $l/R$ are found, one for each value of $\alpha$. The two posteriors also differ.
There is shift with the maximum a posteriori estimate of the posterior for $\alpha=\pi/2$ being a factor $\pi/2$ larger than the one corresponding to the posterior for $\alpha=1$ (recall our previous discussion after Eq.~[\ref{G1}]). Again, the solution is not unique because it depends on the adopted value of $\alpha$. Even for a given $\alpha$, some values of $l/R$ are more plausible than others.

The dispersion of the obtained posteriors depends on the assumed uncertainty on the damping ratio. Decreasing the observational error leads to narrower posteriors. In the limit of very small uncertainty in the damping ratio, $\sigma\rightarrow0$, the posteriors narrow down around the discrete estimates obtained from algebraic inversion of Eq.~(\ref{tauD2}). These discrete estimates for $\zeta$ are $\zeta(\tau_{\rm d}/ P=2, l/R=0.6,\alpha=1)=2.02$ and $\zeta(\tau_{\rm d}/ P=2, l/R=0.6,\alpha=\pi/2)=3.22$. For the transverse inhomogeneity length scale, the discrete estimates are $l/R(\tau_{\rm d}/ P=2, \zeta=3,\alpha=1)= 0.41$ and  $l/R(\tau_{\rm d}/ P=2, \zeta=3,\alpha=\pi/2)= 0.63$. Notice the factor $\pi/2$ between the two analytical estimates for $l/R$. The analytical estimates are shown as vertical solid lines in Figs.~\ref{bayestwoalphas}c and d. 

The solutions obtained by adopting a given value of $\zeta$ or $l/R$ completely remove the uncertainty on the unknown for which a fixed value was taken\footnote{Indeed, taking cuts on a higher dimensional posterior is equivalent to solving the inverse problem using the lower dimensional direct model resulting from considering one of the parameters is a constant.}. When little is known about the fixed parameter, this might lead to the false impression that we know more than what we really know for both of them. One of the advantages of the Bayesian approach is that one must specify explicitly what one is willing to accept about the parameters in the problem. The mechanism to do so is by specifying prior probabilities. When some information on one of the parameters is available, one can incorporate this knowledge by using more informative priors, for example a Gaussian function \citep{arregui11b}. If no additional information is present (or is willing to be accepted), the proper way to infer $\zeta$ (alternatively $l/R$) is to collapse all the information in Figs.~\ref{bayestwoalphas}a and b by marginalising the joint posterior over $l/R$ (alternatively $\zeta$). The resulting marginal posteriors are shown in Figs.~\ref{bayestwoalphas}e and f. They summarise all we can say about the two parameters of interest. The resulting posteriors for the two values of $\alpha$ differ. The difference is more marked for the posterior for $l/R$ (Fig.~\ref{bayestwoalphas}e) than for $\zeta$ (Fig.~\ref{bayestwoalphas}f). Once more,  even in we consider a fixed value for $\alpha$, the solution is not unique and the obtained posteriors indicate the different levels of plausibility of all considered alternative solutions. Also, notice that the maximum a posteriori estimates for the posteriors in Figs.~\ref{bayestwoalphas}e and f, do not match the discrete estimates in Figs.~\ref{bayestwoalphas}c and d. The reason is that the full marginal posteriors for a given parameter take into account all the uncertainty on the other parameter, uncertainty that was removed by considering the fixed parameter as certain. In other words, the discrete algebraic estimate or the corresponding Bayesian cut of the joint posteriors are not good estimators of the marginalised posteriors. 

Figures~\ref{bayestwoalphas}e and f show how different the solution to the inverse problem can be when considering two alternative density profiles at the non-uniform layer. They were computed for a fixed value of the error on the observable damping ratio.We next computed the same marginal posteriors for $\zeta$ and $l/R$ for the two density models and for another damping ratio,  $r=\tau_{\rm D}/ P=6$ representing week damping. The results for both damping ratios ($r=2$ and $r=6$) are compared in Figure~\ref{diffwithdampingrate} (we note that panels~\ref{diffwithdampingrate}a and b are the same as panels~\ref{bayestwoalphas}e and f). The new computations show that the weaker the damping is the less the adopted density model ($\alpha$) affects the inference results and the more closer the posteriors for the two values of $\alpha$ are.  This is in agreement with the conclusion reached by \cite{arregui15c} who computed Bayes factors trying to gather evidence in favour/against a particular model. \cite{arregui15c} found that evidence in favour of a given model is easier to obtain in strong damping regimes. Our results show that this is because the posteriors are more differentiated for stronger damping. In any case, the solution to the inverse problem still depends on the adopted value of $\alpha$ and even for a fixed value of $\alpha$ the solution is never unique.

\begin{figure*}[t]
\center
\includegraphics[width = 0.49\textwidth]{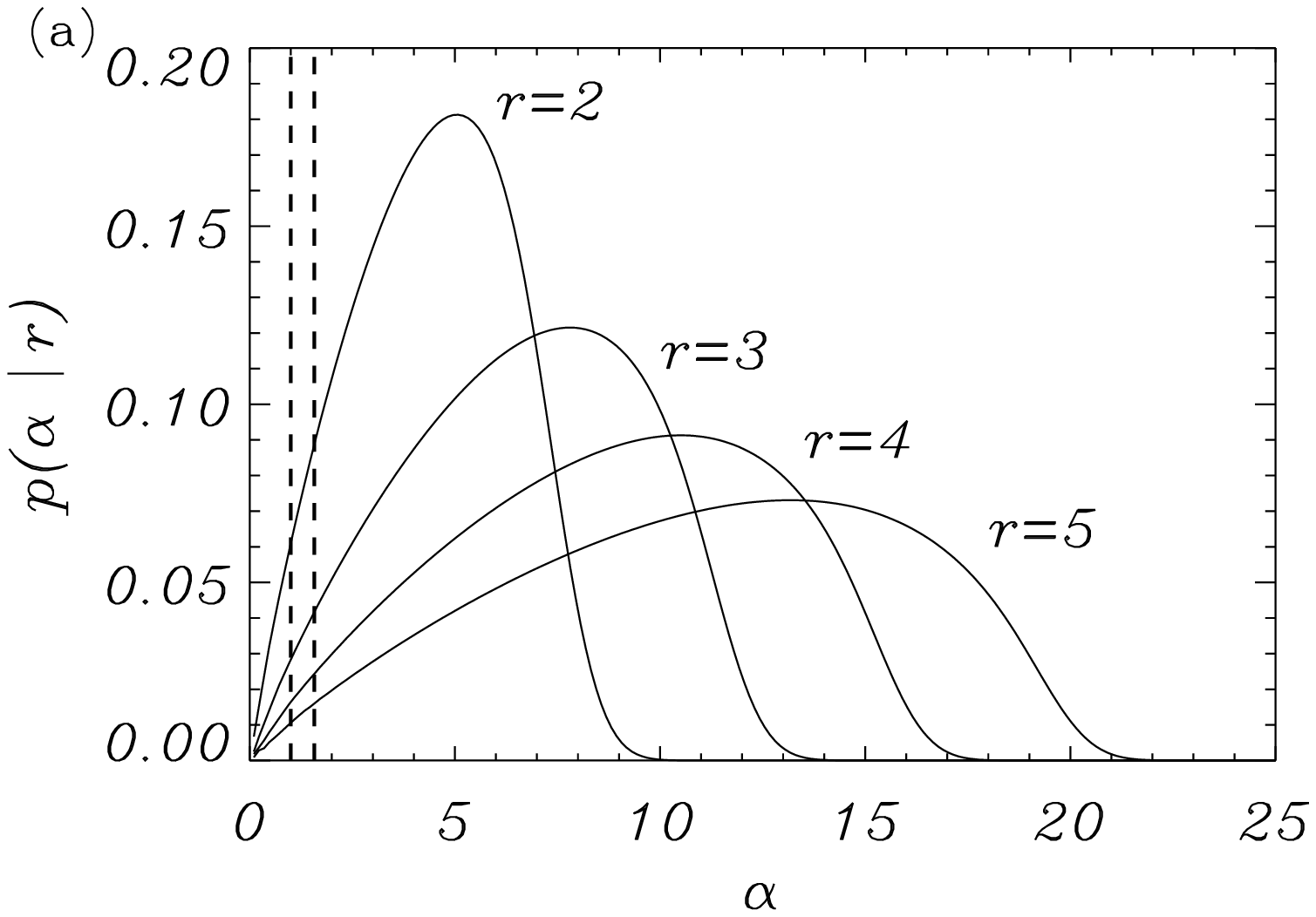} 
\includegraphics[width = 0.49\textwidth]{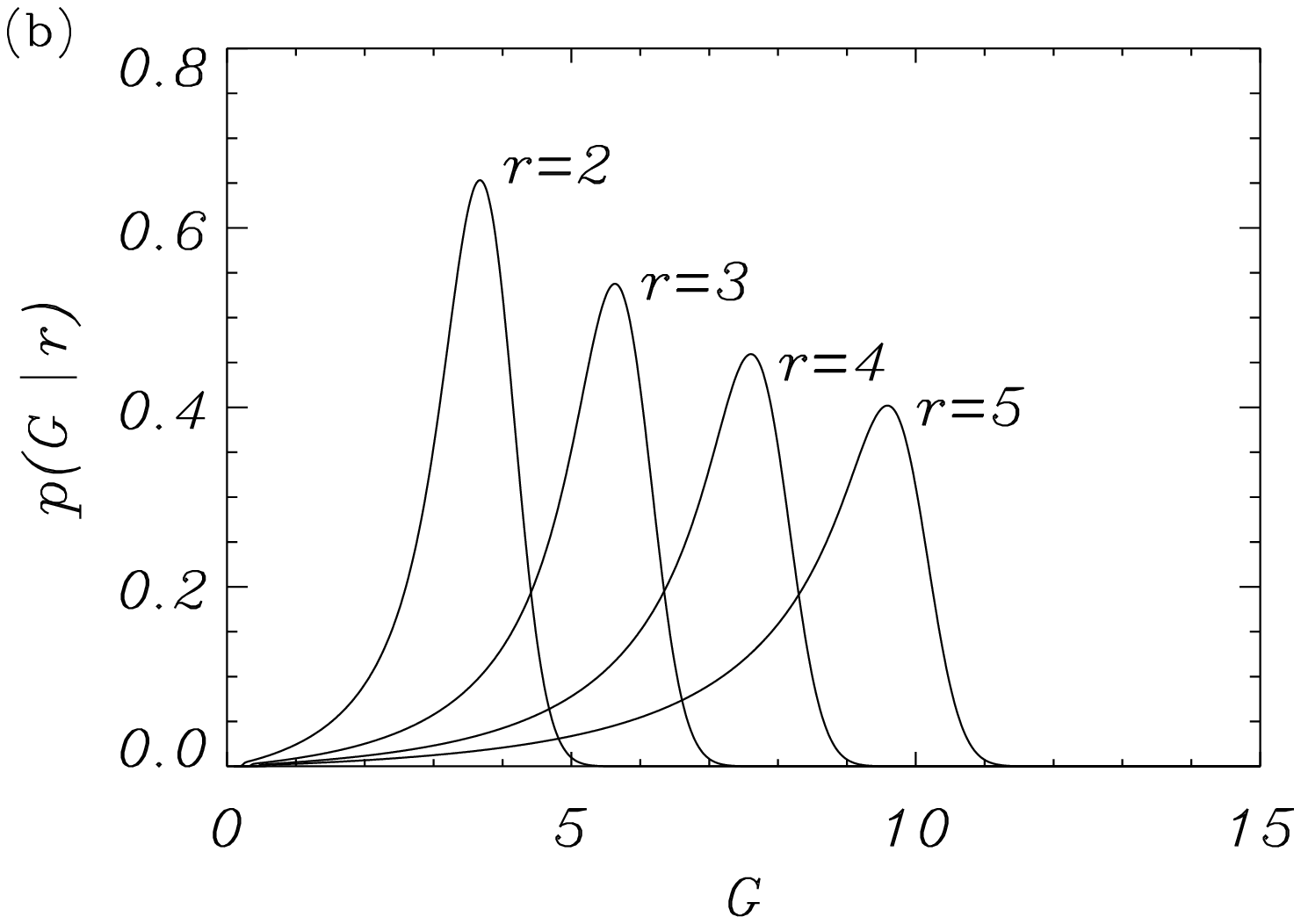} \
\caption{Marginal posteriors for (a) $\alpha$ and (b) G for different values of the damping ratio $r=\tau_{\rm d}/P$ obtained by inversion of Eqs.~\ref{tauD2} and (\ref{tauD1}) and by considering them as additional parameters together with $\zeta$ and $l/R$. In (a) the vertical dashed lines indicate the values $\alpha=1$ (linear density model) and $\alpha=\pi/2$ (sinusoidal density model). The mean and error at 68\% credible intervals for the posteriors in (a) are: $\alpha=4.6^{+1.9}_{-2.3}$ for $r=2$; $\alpha=6.9^{+2.8}_{-3.3}$ for $r=3$; $\alpha=9.2^{+3.8}_{-4.4}$ for $r=4$; and $\alpha=11.5^{+4.8}_{-5.5}$ for $r=5$. The mean and error at 68\% credible intervals for the posteriors in (b) are: $G=3.5^{+0.6}_{-0.8}$ for $r=2$; $G=5.3^{+0.7}_{-1.2}$ for $r=3$; $G=7.1^{+0.8}_{-1.7}$ for $r=4$; and $G=8.9^{+0.9}_{-2.1}$ for $r=5$. In all computations the error in damping ratio is fixed to $\sigma=0.2$. \label{alphag}}
\end{figure*}




\subsection{Inference of $\alpha$ and $G$}

We can now move one step further and consider the inversion of the direct problems given by Eqs.~(\ref{tauD1}) and (\ref{tauD2}) by considering $G$ or $\alpha$ as parameters. Instead of considering that $\alpha$ takes on discrete values arising from the adoption of particular theoretical density models, for example $\alpha = 1$ for a linear profile; $\alpha=\pi/2$ for a sinusoidal profile ; or even $\alpha= \sqrt{2}$ for a parabolic profile as in \cite{arregui15c}, we can let $\alpha$ and also $G$ to be continuously varying parameters.  Notice that from the Bayesian perspective adopting a particular density model and hence fixing the value of $\alpha$ (or $G$) translates into considering that the uncertainty of $\alpha$ or $G$ is zero, a strong model assumption. The Bayesian approach enables to relax this assumption by considering $\alpha$ or $G$ as parameters and by explicitly imposing priors on them. 

We now have two possible problems with one observable, $r=\tau_{\rm d}/P$. In the first case, with direct problem Eq.~(\ref{tauD1}) and considering $G$ as a new parameter, there are two explicit unknowns  ($G$ and $\zeta$) with the additional $l/R$ contained in $G$. In the second case, with direct problem Eq.~(\ref{tauD2}) and considering $\alpha$ as a new parameter, there are three unknowns ($\alpha$, $\zeta$, $l/R$).  Solving these two problems can be seen as performing a model comparison application between alternative density profiles with the difference that they are now characterised by the two continuous parameters $\alpha$ and $G$. In our example applications, we considered $\alpha\in[0.1,25]$ and $G\in[0.1,15]$.  The computed full posteriors $p$($\{\alpha$, $\zeta$, $l/R\}$| $D$) or alternatively $p$($\{G$, $\zeta\}$| $D$) are then marginalised by integrating with respect to the remaining parameters, $\zeta\in[1.1-10]$ and $l/R\in[0.01-2]$ in the first case and with respect to $\zeta$ only in the second case.

The results are shown in Fig.~\ref{alphag}a for the first problem giving the marginal posterior of $\alpha$ and in Fig.~\ref{alphag}b for the second problem giving the marginal posterior of $G$. The different curves correspond to different values of the observable damping ratio, trying to cover week and strong damping cases. The first result is that regardless of whether the inference of $\alpha$ or $G$ is attempted, both can be inferred and show well-constrained probability density functions. In both cases too, the distributions shift towards smaller values of the inferred parameter for stronger damping regimes. 

Concerning $\alpha$, in Fig.~\ref{alphag}a we over-plotted vertical lines indicating the values of $\alpha$ for the linear and sinusoidal density models. For all the curves shown in Fig.~\ref{alphag}a, the maximum a posteriori estimates of $\alpha$ are well above the one corresponding to the values $\alpha=1$ or $\alpha=\pi/2$. This implies that the relative plausibility of the linear and sinusoidal models is lower than that of other possible models with larger $\alpha$. 

Regarding $G$, we mentioned before that Eq.~(\ref{G1}) implies that there are infinitely many couples ($\alpha$, $l/R$) that produce a given damping ratio. The curves in Fig.~\ref{alphag}b simply show how the plausibility of those infinite number of values of $G$ is distributed over the considered range. 

So far we have been unable to give an interpretation to the large values of $\alpha$ or $G$ at which their posteriors show the largest probability values. 

\section{Summary and conclusions}

The classic seismological problem using observations and theoretical expressions for the periods and damping times of transverse standing magnetohydrodynamic (MHD) waves in coronal loops deals with performing algebraic inversions using discrete values for observables and parameters. In general, only a small number of characteristic quantities of the equilibrium profiles can be determined, hence these problems are better referred to as reduced seismological problems.  In particular, there is no reason to adopt a priori  a linear or sinusoidal variation of cross-field density. Even if a particular density model is adopted, the reduced seismological problem does not allow a unique solution.  In particular, infinitely many couples of ($\alpha$, $l/R$) produce the same value for the damping ratio.

These statements were demonstrated by analysing the direct and inverse problems for transverse kink waves in cylindrical straight equilibrium models damped by resonant absorption. The mathematical formulation was generalised by deriving an expression for the damping ratio as a function of a dimensionless quantity, $G$, that relates the local variation of density at the resonant position to the global variation of the density over the mean radius of the cylinder. All the information on the variation of density is collapsed into this quantity. From the point of coronal seismology this result implies that the original seismological problem has infinitely many solutions.

There is nothing wrong with infinitely many solutions. We only need a tool to quantify their relative plausibility. We further support our arguments by considering the inversion problem in the Bayesian framework. Here the solution can never be unique because the inversion procedure involves the computation of posterior density functions that quantify the level of plausibility between alternative parameters/models.  In our example application, the solution to a problem with a fixed density profile ($\alpha$) with two parameters ($\zeta$ and $l/R$) and one observable ($\tau_{\rm d}/P$) leads to a two-dimensional posterior distribution for different combinations of the two parameters. When comparing the inference results for two density models, these joint distributions; their cuts along fixed values of one of the parameters and the marginal posteriors of the parameters differ. 

Moving one step further, we performed seismology with exponential damping by considering the quantities $\alpha$ and $G$ as additional parameters. The inference results show that well constrained posterior probability density functions can be obtained for them. Of particular relevance is the inference of $G$, which contains all the relevant information on the variation of density in the non-uniform layer. We found that the posteriors for both $\alpha$ and $G$ depend on the damping ratio and computed them for different values of this observable.

In addition to offering support to our arguments, our Bayesian computations offer insight into the relationship between algebraic solutions using discrete values of observables and parameters and probabilistic plausibility distributions.
Seismological inference is not the only source of information about the transverse density profile. Additional information can be extracted from the EUV transverse intensity profile observed by EUV imagers such as SDO/AIA.  For a particular event, \cite{pascoe18} found that linear and sinusoidal models are the most consistent with the observed EUV intensity profile. Synthetic and observed EUV profiles could be incorporated into a Bayesian analysis to obtain further information.

%
\small  
\section*{Acknowledgments}   
We acknowledge support by the Spanish Ministry of Economy and Competitiveness (MINECO) through projects 
AYA2014-55456-P (Bayesian Analysis of the Solar Corona) and AYA2014-60476-P (Solar Magnetometry in the Era of Large Solar Telescopes) and FEDER funds.

\end{document}